\documentclass[aps,twocolumn]{revtex4}
\usepackage{epsfig}
\usepackage{graphicx}
\usepackage{amsmath,amssymb,amsfonts}
\usepackage{array}
\usepackage{url}
\usepackage{hyperref}
\usepackage{multirow}
\usepackage{float}
\usepackage{lineno}
\usepackage{xspace}
\usepackage[usenames,dvipsnames]{color}

\newcommand{\bef}{\begin{figure}}
\newcommand{\eef}{\end{figure}}
\newcommand{\bc}{\begin{center}}
\newcommand{\ec}{\end{center}}

\newcommand{\be}{\begin{equation}}
\newcommand{\ee}{\end{equation}}
\newcommand{\bea}{\begin{eqnarray}}
\newcommand{\eea}{\end{eqnarray}}

\def\ba{\begin{eqnarray}}
\def\ea{\end{eqnarray}}
\begin{document}
\title{Bose-Einstein Condensation and Dissipative Dynamics in a Relativistic Pion Gas}



\author{Kshitish Kumar Pradhan}
\author{Dushmanta Sahu}
\author{Captain R. Singh}
\author{Raghunath Sahoo\footnote{Corresponding author Email: Raghunath.Sahoo@cern.ch}}
\affiliation{Department of Physics, Indian Institute of Technology Indore, Simrol, Indore 453552, India}

\begin{abstract}

Pion condensation in ultra-relativistic collisions presents a compelling theoretical phenomenon with significant implications for the dynamics of hadronic matter. Various theoretical frameworks offer insight into the nature of high-temperature Bose-Einstein condensation (BEC). The present study investigates the dissipative behavior of a relativistic pion gas undergoing Bose-Einstein condensation (BEC) in ultra-relativistic heavy-ion collisions. Further, we obtain viscosity ($\eta$), bulk viscosity ($\zeta$), and speed of sound ($c_s$)  by employing the Boltzmann transport equation with the relaxation time approximation. Findings show a substantial drop in $\eta/s$ and $\zeta/s$ with the fractional increase in condensation. This effect is becoming more evident in larger systems approaching the thermodynamic limit. Alongside the reduction in viscosities, the speed of sound also decreases with increasing condensation, indicating a softening of the equation of state. The analysis of finite-size effects reveals that larger systems exhibit more pronounced signatures of BEC. These results suggest that pion condensation can influence the hydrodynamic evolution of the hadronic phase in heavy-ion collisions, with consequential implications for interpreting collective flow observables and the underlying equation of state.

\end{abstract}
\date{\today}
\maketitle

\section{Introduction}
\label{intro}

The formation of hot and dense matter in ultra-relativistic collisions at the Relativistic Heavy Ion Collider (RHIC) at Brokhaven National Laboratory (BNL) and the Large Hadron Collider (LHC) at CERN leads to numerous exciting consequences. The lightest hadron, pions, are the most abundant particles in the final state of such collisions. Pions, being mesons (bosons), have an integer spin and thus follow Bose-Einstein (BE) statistics, named after Satyendra Nath Bose and Albert Einstein. Following the ideas of Bose's 1924 paper on Planck's law and light quantum hypothesis \cite{Bose}, Einstein extended them to massive particles by formulating the quantum theory of ideal gas in two series, first in 1924 \cite{Einstein} and the second one in 1925 \cite{Einstein_02}. In the later work, Einstein argued that a system of bosons has a unique property to condense in the ground energy state at temperatures $T\sim0^{o}$ K and when their momentum becomes almost zero~\cite{Bose, Einstein}. This phenomenon is now known as Bose-Einstein condensation (BEC). Since bosons have symmetric wave functions, a large number of bosons can occupy the ground state where their wave functions can interfere with each other. This effect is then observed macroscopically. This phenomenon eluded physicists for a long time, as achieving the conditions required for BEC formation is tricky. However, in 1995, BEC in cold Rubidium atoms was observed experimentally, confirming the decades-long predictions to be correct ~\cite{Anderson:1995gf}. Since then, BEC has been one of the most exciting research topics in various fields, from condensed matter to astrophysics and high-energy physics ~\cite{Begun:2006gj,Begun:2008hq,Strinati:2018wdg,Nozieres:1985zz,Funaki:2008gb,Chavanis:2011cz,Mishustin:2019otg,Padilla:2019fju}.\\

High pion density in the hadronic medium offers a possible condition for the onset of the Bose-Einstein condensation. Such a behavior has been explored under a chemical non-equilibrium model where the pion abundances are characterized by non-zero chemical potential~\cite{Letessier:1998sz}, which might indicate an onset for BEC at the LHC energies. Nevertheless, due to their high density in the hadronic phase of the collision, one may expect the formation of the BEC of pions. However, the question arises: how can a violently expanding system such as the one formed at the end of an ultra-relativistic collision uphold a BEC-like phenomenon? As it is known, the relative momentum of the particles must be near or equal to zero for the formation of BEC, which is a necessary and sometimes sufficient condition for the BEC phase to exist. In ultra-relativistic collisions, the energy and particle density are so high that the particles can exist very close to each other in the phase space and have zero relative momentum. The small volume of the systems produced in high-energy collisions and the strong interactions involved may provide such an environment for BEC to exist even at high temperatures. The possibility of high-temperature BEC in ultra-relativistic collisions is fascinating and has been studied previously ~~\cite{Begun:2015ifa,Begun:2010ec,Begun:2008hq,Deb:2021gzg}. Moreover, interactions between pions as defined by excluded volume and mean-field approximation have also been considered to explore the onset of BEC in a pion gas~\cite{Savchuk:2020yxc}. In addition, in a system of interacting pions, the location of a critical point has also been explored ~\cite{Kuznietsov:2021lax}. The available literature has also reported that pions could form condensate in the $pp$ and peripheral Pb-Pb collisions at the LHC ~~\cite{Deb:2021gzg}. Though, the formation of BEC in ultra-relativistic collisions is still a subject of verification through the experiments, the proposal to find BEC experimentally was based on the concept of an increase in the yields at the lower transverse momentum region of the pion $p_{\rm T}$-spectra. However, no prominent signal has been observed yet as the pion spectra get contaminated by the resonance decays, and measurements at very low $p_{\rm T}$ are not yet available.\\

In high-energy heavy-ion collisions, the possibility of coherent pion production has also been considered previously. In recent works at ALICE, analysis of two- and three-pion correlation measurements show almost 23$\%$ coherent fraction in charged pion emission ~\cite{ALICE:2013uhj}. The pion radiation from a BEC is expected to be coherent and thus suppress Bose-Einstein correlations ~~\cite{Fowler:1978us}. Effective suppression of three-pion correlation compared to two-pion correlation at low triplet momentum has been observed, which is consistent with the formation of a Bose-Einstein condensate ~\cite{ALICE:2013uhj}. In addition, the measurement of particle number fluctuation can also act as a signature of BEC \cite{Begun:2008hq}. Thus, it is fascinating to see the outcome of such a phenomenon on the system dynamics.\\

From the elliptic flow measurements at RHIC, a minimum value for shear viscosity to entropy density ratio ($\eta/s$) has been observed ~\cite{STAR:2005gfr}, approaching the KSS bound limit ($\eta/s = 1/4\pi$) estimated from the AdS/CFT calculations ~\cite{Kovtun:2004de}. The $\eta/s$, which shows minima near the QCD critical phase transition, is crucial in understanding the behavior of the matter formed in ultra-relativistic energies. Many studies have been done to estimate shear viscosity in hot QCD matter and in hadronic matter. However, unlike $\eta/s$, the bulk viscosity to entropy density ratio ($\zeta/s$) is poorly explored at very high temperatures, where conformal symmetry in the QCD matter holds; $\zeta/s \to$ 0. Nevertheless, at the QCD critical temperature, the conformal symmetry breaks, which results in the observation of a peak in the trace anomaly ($(\epsilon - 3P)/T^{4}$). Thus, one expects similar behavior in $\zeta/s$ near the critical temperature. The dissipative properties, such as shear viscosity and bulk viscosity, are essential to study the behavior of the matter formed in ultra-relativistic collisions. In Ref.~\cite{Chen:2018mwr}, the authors have taken a gluon plasma and studied the effect of BEC on the shear viscosity of the system with the Green-Kubo method. They observed the shear viscosity to be decreasing under BEC. Given this, it would be interesting to have a study on the pion gas and the effect of pion condensation on the viscosity. In this work, the BEC properties of pions are studied, and further estimate is made on the shear ($\eta$) and bulk viscosity ($\zeta$) of a pion gas under condensation. The Boltzmann transport equation is employed under relaxation time approximation to study the viscosities of a pion gas under BEC. The entropy density ($s$), specific heat ($c_{v}$), and the speed of sound ($c_{s}$) are also studied in the pion gas under the effect of BEC to understand the nature of the medium. This discussion is justified for two main reasons. Firstly, pions are the most abundant particle species in the final state of the ultra-relativistic collision. Secondly, as suggested by the chiral perturbation theory earlier ~\cite{Son:2000xc}, recent results from lattice QCD support the existence of pion BEC at a finite isospin chemical potential ~\cite{Brandt:2017oyy}. Moreover, studying pion BEC has many applications apart from the systems formed in ultra-relativistic collisions, such as hypothetical pion stars~\cite{Brandt:2018bwq,Andersen:2018nzq} and cosmic trajectory in the early universe~\cite{Brandt:2017oyy,Abuki:2009hx}.\\

This article is organized as follows:  a detailed formulation of pion BEC, its thermodynamics, and subsequently the shear and bulk viscosity and the speed of sound estimation for a hot pion gas under BEC formulation are characterized in Sec.~\ref{formulation}. In Sec.~\ref{res}, a detailed discussion is carried out on the obtained results and their implications. Finally, in Sec.~\ref{sum}, the findings are summarized.

\section{Formulation}
\label{formulation}
\subsection{Bose-Einstein condensate}
From basic statistical mechanics, the Bose-Einstein distribution function is given by~\cite{KHuang},
\begin{equation}
f = \left[{\rm exp}\left(\frac{E - \mu}{T}\right) - 1\right]^{-1}
\label{eq1}
\end{equation}
Here, $E = \sqrt{p^{2}+m^{2}}$ is the energy of the particle, $p$ is the momentum, $m$ is the mass of the particle, $\mu$ is the chemical potential, and $T$ is the temperature of the system. The particle multiplicities can be calculated from the equation in natural units ($c = \hbar = 1$) as~\cite{Begun:2015ifa},
\begin{equation}
\label{eq2}
N =  g\int\frac{d^3xd^3p}{(2\pi)^3}  \left[{\rm exp} \left(\frac{\sqrt{p^2 + m^2} - \mu}{T}\right) - 1\right]^{-1}  \nonumber
\end{equation}
\begin{equation}
\simeq  \frac{gV}{(2\pi)^3} \int d^{3}p\left[{\rm exp} \left(\frac{\sqrt{ p^2 + m^2} - \mu}{T}\right) - 1\right]^{-1}
\label{eq3}
\end{equation}
where $g$ is the degeneracy of the particle, $h$ is Planck's constant, and $V$ is the volume of the system. The total number density ($n_{\rm tot}$) can therefore be written as
\begin{equation}
\label{thlim}
\begin{split}
    n_{\rm tot} &= \frac{g}{(2\pi)^3} \int \frac{d^{3}p}{{\rm exp} \left(\frac{\sqrt{p^2 + m^2} - \mu}{T}\right) - 1} \\
    &\approx \frac{gTm^2}{2\pi^2}\sum_{r=1}^{\infty}\frac{1}{r}K_2(rm/T)\exp(r\mu/T),
\end{split}
\end{equation}
where $K_2$ is the modified Bessel function and $r$ is a positive integer. In the thermodynamic limit (for $V\to\infty$), the sum over momentum states is transformed into the momentum integral. This works perfectly fine when $\mu < m$. However, when $\mu \to m$, something interesting happens. The integrand in the first line of Eq.~(\ref{thlim}) stays constant for $p$ $\to$ 0. This is due to the fact that the singularity of the denominator is cancelled by the integration measure $d^{3}\it {p}$. However, summing over quantum levels, the first term becomes infinite for $\mu = m$ at $p = 0$, i.e.,
\begin{equation}
\label{ncinfty}
 n_{\rm cond} \simeq  \frac {g}{ {V[\rm exp} (\frac{m - \mu}{T}) - 1]} \to \infty ~\rm {for} ~\mu \to m
  \end{equation}

 Thus, when $\mu$ approaches from lower values to $m$, the ground state starts populating and becomes too large at $\mu = m$. This marks the onset of BEC, and the integration of Eq.~(\ref{eq2}) should begin for $p >$ 0 while keeping the summation over low momentum. In the thermodynamic limit ~\cite{Begun:2008hq}, one can write Eq. ~(\ref{eq2}) with separate terms for $p$ = 0 and $p$ $>$ 0, as:
\begin{equation}
\begin{split}
\label{n_finite}
    &n_{\rm tot} \simeq  \frac {g}{ V[{\rm exp} (\frac{m - \mu}{T}) - 1]} + \int \frac{d^3p}{(2\pi)^3} \frac {g}{{\rm exp} \bigg(\frac{\sqrt{p^2 + m^2} - \mu}{T}\bigg) - 1}\\  
    &\Rightarrow n_{\rm tot} = n_{\rm cond} + n_{\rm ex}.
\end{split}
\end{equation}
Here, $n_{\rm cond}$ and $n_{\rm ex}$ are the number densities in the condensate and in the excited states, respectively. Here we are focusing on a statistical analysis of the Grand Canonical Ensemble of pion gas, which takes into account the total pion density after chemical freeze-out. This includes contributions from the decays of heavier resonances, allowing for meaningful comparisons with experimental analyses. One can observe that since the condensate density is given by $n_{\rm cond} = (g/V)[\exp{(m-\mu)/T}-1]^{-1}$, for $\mu\to m$, a finite number of particles appear in the condensed state, i.e., $n_{\rm cond}>0$, and thus, the difference between the ground state energy and chemical potential is given by \cite{chapter3Y},
\begin{equation}
\label{m_mu}
    m-\mu = T \ln\Big(1+\frac{g}{n_{\rm cond}V}\Big).
\end{equation}
However, this condition breaks down when $n_{\rm cond}$ = 0, then one has $n_{\rm ex}$ = $n_{\rm tot}$. Therefore, the transition from a non-condensed state to a BEC state implies a transition from a zero to a positive value of $n_{\rm cond}$. Hence, the critical temperature $T_c$ of BEC can be defined as the lowest temperature at which the excited states consist of the maximum number of particles \cite{chapter3Y, megias2022, chapter12SM}, i.e.,
\begin{equation}
    n_{critical} = n_{\rm ex}(T=T_c, \mu = m) = n_{\rm tot}
\end{equation}
Below the critical temperature ($T< T_c$), the particle density in the excited states becomes smaller than that at $T_c$, and therefore, the excess particles must be at the ground state. The condensate fraction (ratio of condensate number density to total number density) is then given by
\begin{equation}
    \label{cfraction}
    \frac{n_{\rm cond}}{n_{\rm tot}} = 1 - \frac{n_{\rm ex}}{n_{\rm tot}} = 1-\frac{n_{\rm ex}(T, \mu = m)}{n_{\rm ex} (T_c, \mu = m)}.
\end{equation}
  \begin{figure*}[ht!]
\begin{center}
\includegraphics[scale = 0.44]{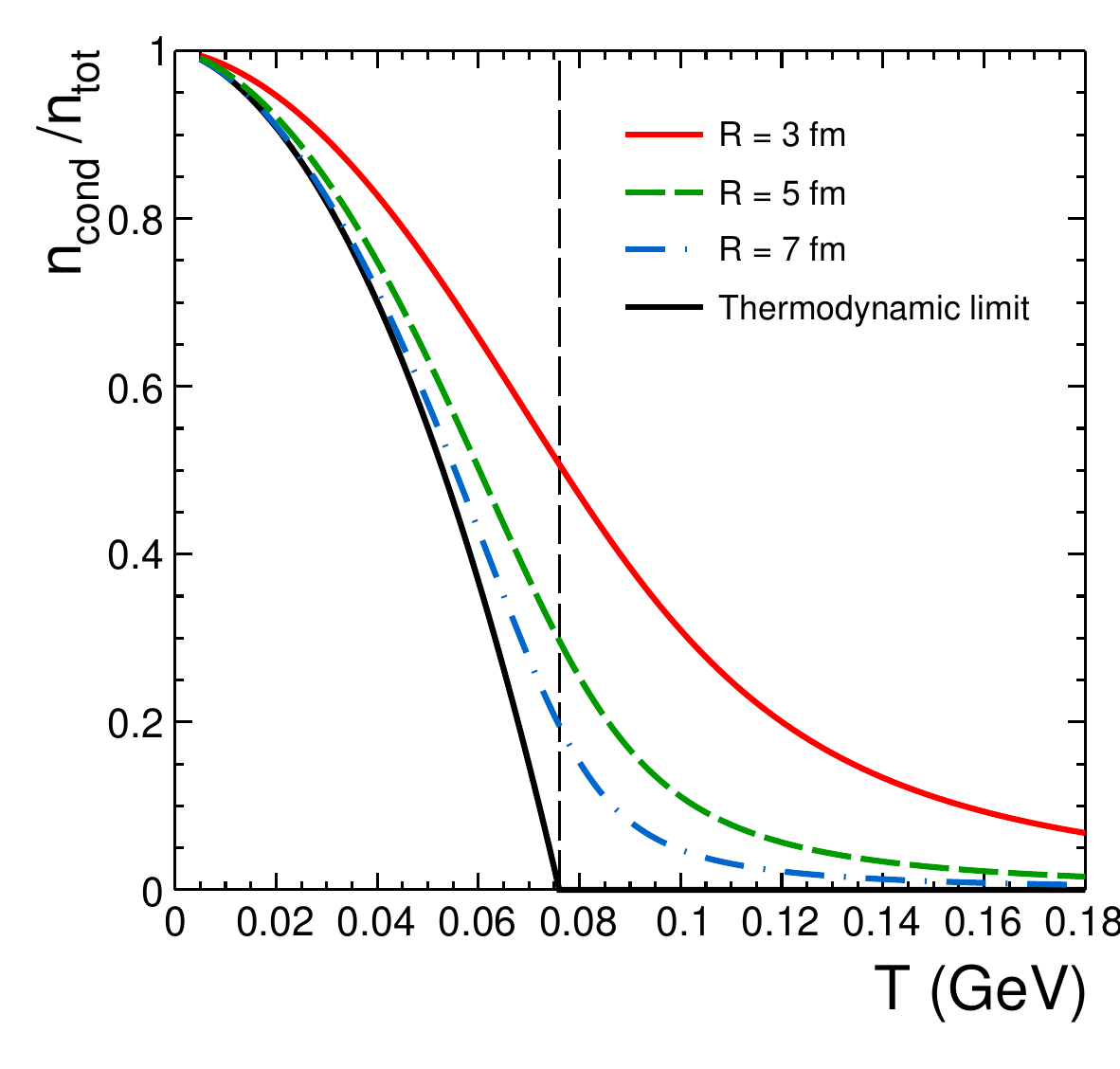}
\includegraphics[scale = 0.44]{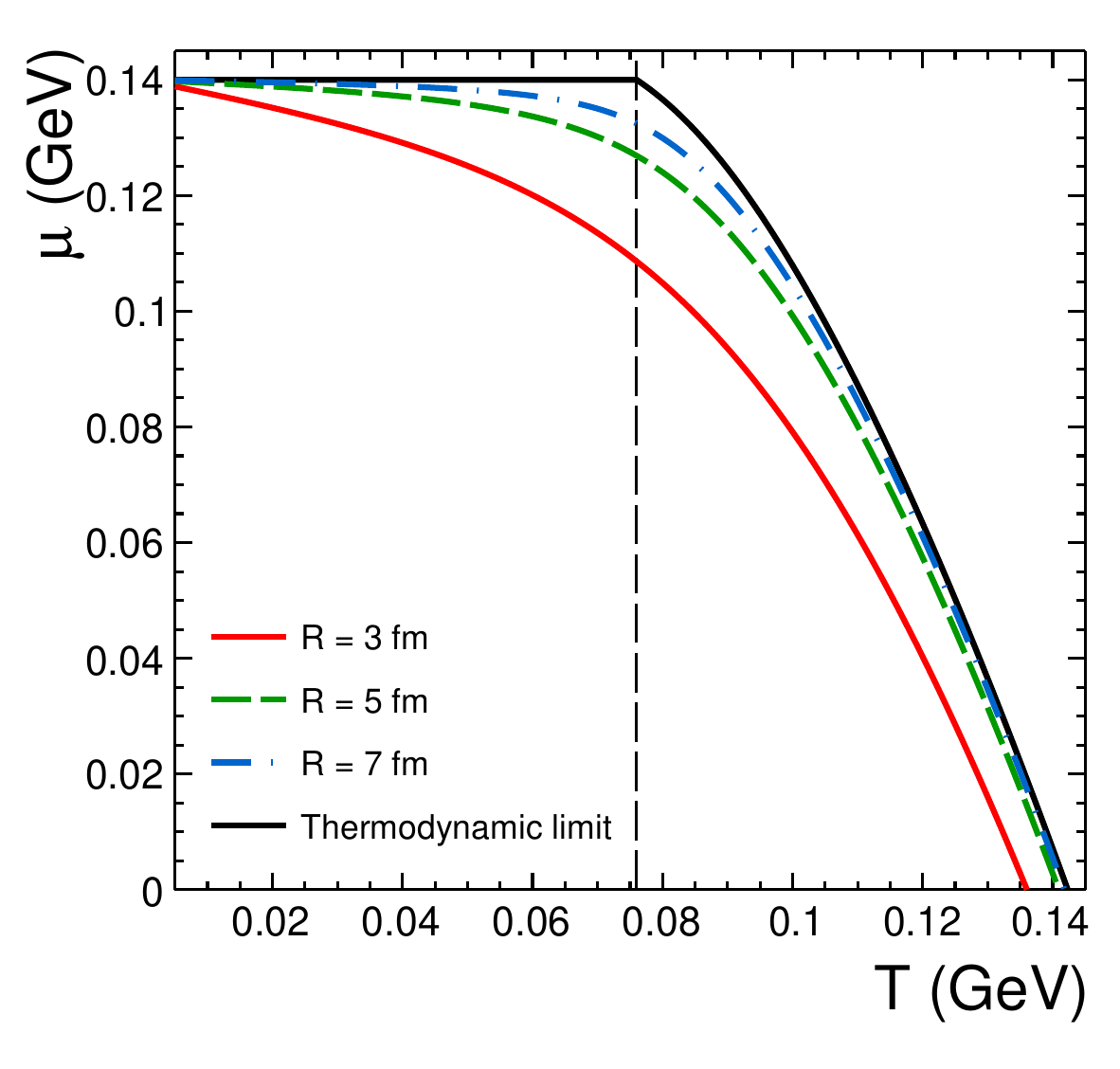}
\caption{(Colour Online) The left panel shows the condensate fraction as a function of temperature, and the right panel shows the temperature dependence of chemical potential, $\mu$, in the case of the thermodynamic limit as well as for different systems with size $R$ = 3, 5, and 7 fm. This is obtained by considering a fixed density of $n_{\rm tot}$ = 0.1 fm$^{-3}$ which corresponds to a critical temperature, $T_c$ = 0.076 GeV.}

\label{fig1}
\end{center}
\end{figure*}
As $T\to0$, the condensate fraction ($n_{\rm cond}/n_{\rm tot}$) approaches unity, implying all the particles are condensed. One must consider the thermodynamic limit for a BEC phase transition to be observed, and for this, one needs to start with a finite volume system \cite{Begun:2008hq}. In this work, a pion gas is considered. Neglecting the small mass difference among the charged and neutral pions, the pion mass is taken as $m\simeq$ 0.14 GeV for simplicity, with degeneracy $g$ to be 3. In the left panel of Fig.~\ref{fig1}, the condensate fraction is shown as a function of temperature in a pion gas having fixed number density ($n_{\rm tot}$) = 0.1 fm$^{-3}$. A study on the dependence of number density has already been done in Ref.~\cite{Begun:2008hq}. The black solid line represents the condensate fraction in the thermodynamic limit, for which the critical temperature is obtained at $T = T_C = $ 76 MeV, obtained using Eq.~(\ref{thlim}). This is in line with what was observed in earlier studies~\cite{Begun:2008hq}. In the current work, the focus is on the finite size effects on the BEC, as these kinds of studies may help to identify any signatures of BEC in experiments, for example, a probable pion BEC fluctuation as proposed in~\cite{Begun:2007yk, Begun:2008hq}. Therefore, to examine the finite size effects, we model the system as a sphere of radius R, which determines the volume of the system by $V=(4/3) \pi R^3$. Three different system sizes with radius $R$ = 3, 5, and 7 fm are considered. One can estimate the total number of particles when total density is kept fixed (here $n_{\rm tot}$ = 0.1 fm$^{-3}$). Therefore, a large system implies a larger radius and hence an increase in total particle number. This is employed to study the impact of finite system-size effects \cite{Scaria:2022yrz} on the properties of a BEC. For the finite volume cases, Eq.~(\ref{n_finite}) is used to estimate the condensate fraction. The system radius enters the thermodynamic expressions through this volume dependence, affecting the population of the condensate and the excited states. Here, one must be careful with the treatment of the ground state. In finite systems, the quantization of momentum levels implies that the lowest momentum state is not at $p=0$, but is determined by the boundary conditions of the system. The condensate then occupies the lowest allowed momentum state, which still exhibits an enhanced occupation, though a strict divergence of the condensate population will not be there, and the BEC transition is smoothed in comparison to the case of the thermodynamic limit ($R\to\infty$). In the current work, we found that even for the smallest system size ($R$ = 3 fm), the change in the ground state energy is smaller than that of the thermodynamic limit, which has negligible effects on the results presented here. Again, due to discretized momentum spectrum because of the boundary conditions, replacing momentum integrals with summation over quantized momentum states would provide a more accurate treatment of the system. While the qualitative trends presented here will be the same, the precise values of the condensate fraction or transport coefficients may vary, though the change is small. A more rigorous treatment incorporating discrete summation could further refine the finite-size effects.

The condensate fraction in a system with different volumes is shown in the left panel of Fig.~\ref{fig1}. It is observed that for a system with finite size, the condensate fraction decreases with $T$, but doesn't vanish at $T_c$ as in the case of the thermodynamic limit. Some particles remain in the condensed state even at $T>T_c$. The fraction becomes asymptotically small with increasing $T$. No sharp transition is observed at $T_c$. The onset of BEC transition is smeared out in finite volume cases. Therefore, the critical temperature ($T_c$) throughout the study is considered as that in the thermodynamic limit. As the volume increases with an increase in $R$, the constraint of fixed number density implies an increase in the number of particles, and the transition approaches the thermodynamic limit where both the particle number and volume of the system approach infinity, keeping the density fixed. Regardless of volume, the condensate fraction approaches unity when $T\to 0$, as expected. Also, the volume dependence ceases when one moves towards very high temperatures. Moreover, since the number density in the condensate and excited state is changing with $T$, the condition of fixed total density $n_{\rm tot}$ leads to a temperature dependence of chemical potential ($\mu\equiv\mu (T)$). At any temperature, the $\mu$ is evaluated so as to keep the total number density fixed. The behaviour of $\mu (T)$ is shown in the right panel of Fig.~\ref{fig1}. Here also, one can observe that with increasing system volume, one approaches the condition of the thermodynamic limit.

\begin{figure}[H]
\centering
\includegraphics[scale = 0.34]{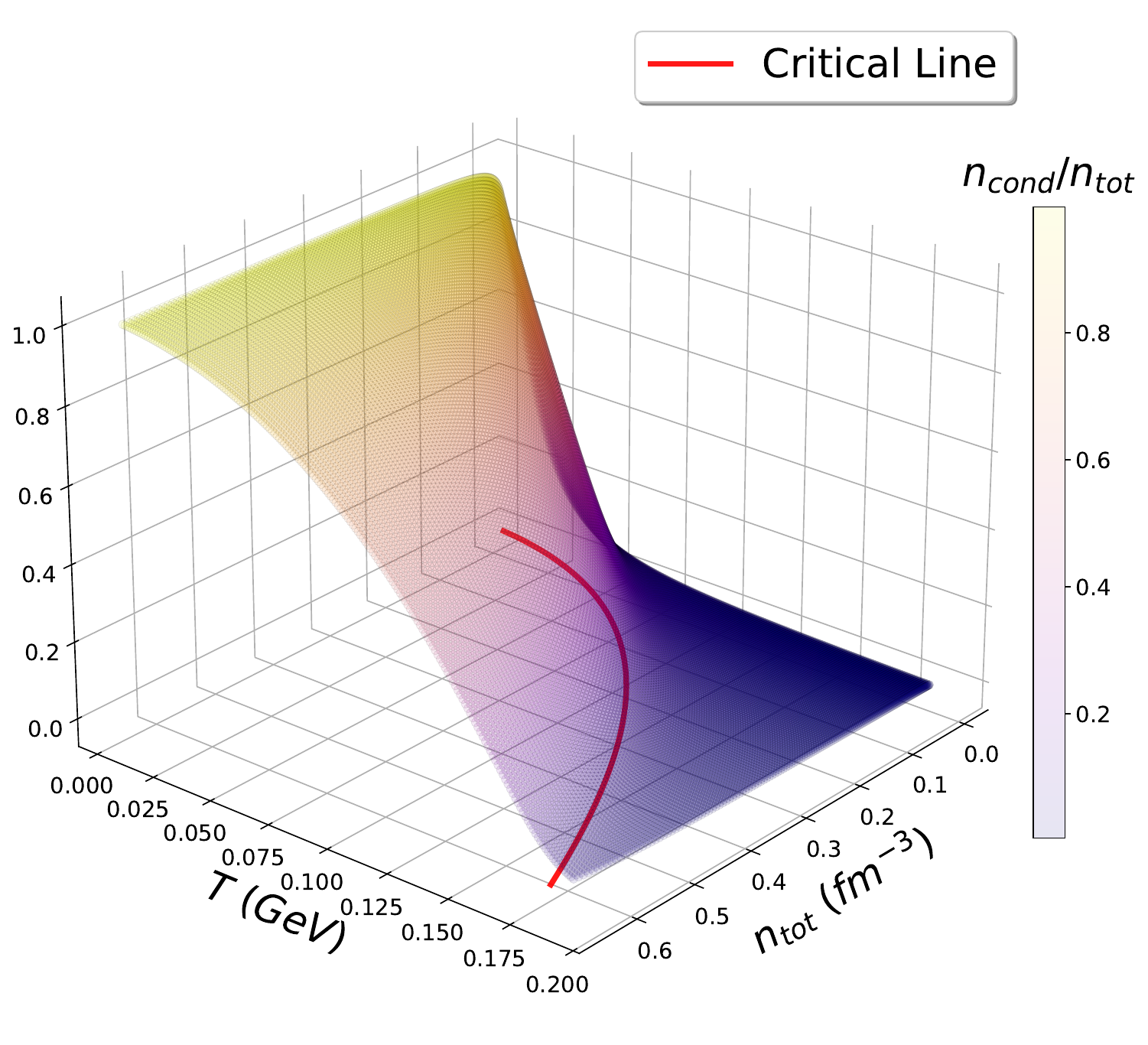}
    \caption{Condensate fraction of pions as a function of temperature for different total number densities. The red curve in the $T-n_{\rm tot}$ plane is the BEC critical line ($T = T_c$) that represents the critical temperature, $T_c$, of the phase transition for different fixed densities. For $T<T_c$, the BEC phase is described by Eq.~(\ref{n_finite}).}
\label{fig2}
\end{figure}

It is to be noted here that the $T_c$ will be different for different cases of total density considered. In Fig.~\ref{fig2}, the condensate fraction as a function of temperature for different values of $n_{\rm tot}$ in the finite volume case (for $R$ = 7 fm) is shown. The red solid curve in the $T-n_{\rm tot}$ plane represents the BEC critical line in the thermodynamic limit. 

\subsection{Shear and bulk viscosity}
The dissipative properties, shear and bulk viscosities, can be estimated from the kinetic theory using the relativistic Boltzmann transport equation (BTE) given by \cite{Gavin}
 \begin{equation}
\frac{\partial f}{\partial t}+ v_{p}^{i}\frac{\partial f_{p}}{\partial x^{i}}=I\{f_{p}\},
\label{boltzeqn}
\end{equation}
where, $\vec v_{p}=\vec p/E_{p}$ is single particle velocity and the collision integral $I\{f_{p}\}$ gives the rate of change of the non-equilibrium distribution function $f_{p}$ due to collisions.
Initially, when the system is slightly away from the equilibrium, subsequent collisions among the constituent pions will try to restore the equilibrium after a certain time. And according to the relaxation time approximation, the collisions bring the system to the local equilibrium exponentially with relaxation time, which is of the order of collision time. Therefore, the collision integral under the RTA takes the form
 \begin{equation}
I\{f_{p}\}\backsimeq -\frac{(f_{p}-f_{p}^{0})}{\tau(E_{p})},
\label{rtapprx}
\end{equation}
where $\tau(E_{p})$ is called relaxation time. The equilibrium distribution function, denoted by $f_{p}^{0}$, is given by 
 \begin{equation}
f_{p}^{0}=\frac{1}{exp\bigg(\frac{E_{p}-\mu}{T}\bigg)\pm1}
\label{distribution}
\end{equation}
where the $\pm$ corresponds to fermions and bosons, respectively. The stress-energy tensor, $T^{\mu\nu}$, for a system can be considered the sum of an ideal part ($T^{\mu\nu}_{0}$) and a dissipative part ($T^{\mu\nu}_{dissi}$) and can be written as,
\begin{equation}
    T^{\mu\nu} = T^{\mu\nu}_{0} + T^{\mu\nu}_{dissi}.
\end{equation}
In a hydrodynamical description of the QCD matter, the shear and bulk viscosities enter the dissipative part ($T_{dissi}^{\mu\nu}$) of the stress-energy tensor, which has the form as
\begin{equation}
    \begin{split}
        \label{dissipative}
        T_{dissi}^{\mu\nu}=-\eta\big(\nabla^\mu u^{\nu}+\nabla^\nu u^\mu\big)+(\zeta-\frac{2}{3}\eta)\Delta^{\mu\nu}\nabla_\alpha u^\alpha
    \end{split}
\end{equation}
In the local Lorentz frame, we have $u^\mu$ = (1, $\vec0$). Then the projection tensor $\Delta^{\mu \nu}$ = $g^{\mu\nu}-u^\mu u^\nu$ = diag(0, 1, 1, 1) and $\nabla^\mu = \Delta^{\mu\alpha}\partial_\alpha = \partial^\mu$. Hence, the above equation with (non-zero) spatial components can be written as \cite{Gavin},
 \begin{equation}
T_{dissi}^{ij}=-\eta\bigg(\frac{\partial u^{i}}{\partial x^{j}}+\frac{\partial u^{j}}{\partial x^{i}}\bigg)-(\zeta-\frac{2}{3}\eta)\frac{\partial u^{i}}{\partial x^{j}}\delta^{ij}
\label{dissi}
\end{equation}

In terms of distribution function, the $T_{dissi}^{ij}$ can be expressed as
 \begin{equation}
T_{dissi}^{ij}= \int \frac{d^{3}p}{(2\pi)^{3}p^{0}}p^{i}p^{j}\delta f_{p}
\label{dissi1}
\end{equation}
where $\delta f_{p}$ is a measure of the deviation of the distribution function from equilibrium and can be obtained from Eq. (\ref{boltzeqn}) and Eq. (\ref{rtapprx}) as,
 \begin{equation}
\delta f_{p}=-\tau(E_{p})\bigg(\frac{\partial f_{p}^{0}}{\partial t}+ v_{p}^{i}\frac{\partial f_{p}^{0}}{\partial x^{i}} \bigg).
\label{deltadistr}
\end{equation}
To calculate the shear viscosity, consider a one-dimensional steady flow of the form $u^{i}=(u_{x}(y),0,0)$ and a space-time independent temperature, then Eq. (\ref{dissi}) reduces to $T^{xy}=-\eta\partial u_{x}/\partial y$. The time derivative in Eq. (\ref{deltadistr}) vanishes for such a flow, and hence from Eq. (\ref{dissi1}) using Eq. (\ref{deltadistr}), one gets,
 \begin{equation}
T^{xy}=\bigg\{-\frac{1}{T}\int\frac{d^{3}p}{(2\pi)^{3}}\tau(E_{p})\bigg(\frac{p_{x}p_{y}}{E_{p}}\bigg)^{2}f_{p}^{0}\bigg\}\frac{\partial u_{x}}{\partial y}.
\end{equation}

Thus, the expression for the coefficient of shear viscosity for a single component of the hadronic matter is finally given by ~\cite{Kadam:2015xsa},
 \begin{equation}
\eta=\frac{1}{15T}\int\frac{d^{3}p}{(2\pi)^{3}}\tau(E_{p})\frac{p^{4}}{E_{p}^{2}}f_{p}^{0}
\end{equation}

Now, the departure of the system from equilibrium when it is compressed uniformly is expressed through the dissipative parameter called bulk viscosity. Taking trace of Eq. (\ref{dissi}), one gets
 \begin{equation}
(T_{dissi})^{ii}=-3\zeta \frac{\partial u^{i}}{\partial x^{i}}
\label{bulkdissi1}
\end{equation}
Also from Eq. (\ref{dissi1}) and Eq. (\ref{deltadistr}),
 \begin{equation}
(T_{dissi})^{ii}=-\int\frac{d^{3}p}{(2\pi)^{3}}\tau(E_{p})\frac{p^{2}}{E_{p}}\bigg(\frac{\partial f_{p}^{0}}{\partial t}+ v_{p}^{i}\frac{\partial f_{p}^{0}}{\partial x^{i}} \bigg)
\label{bulkdissi2}
\end{equation}
In the local rest frame, the energy-momentum conservation implies $\partial_{\mu}T^{\mu\nu}=0$, and together with Eq. (\ref{bulkdissi1}) and Eq. (\ref{bulkdissi2})  one can arrive at the expressin for bulk viscosity~\cite{Gavin} 
 \begin{equation}
\zeta=\frac{1}{T}\int \frac{d^{3}p}{(2\pi)^{3}}\tau(E_{p}) f^{0}_{p}\bigg[E_{p}c_{s}^{2}-\frac{p^{2}}{3E_{p}}\bigg]^{2}
\end{equation}
where $E_{p}^{2}=p^{2}+m_{p}^{2}$, and $c_{s}^{2}=\frac{\partial P}{\partial\varepsilon}$, is the square of the speed of sound at constant baryon density.

The energy-dependent average relaxation time is defined by the expression
 \begin{equation}
 \tau^{-1}(E_{a})=\sum_{bcd}\int\frac{d^{3}p_{b}}{(2\pi)^{3}}\frac{d^{3}p_{c}}{(2\pi)^{3}}\frac{d^{3}p_{d}}{(2\pi)^{3}}W(a,b\rightarrow c,d)f_{b}^{0}
 \label{tdrext}
\end{equation}
where the transition rate $W(a,b\rightarrow c,d)$ is defined by
 \begin{equation}
W(a,b\rightarrow c,d)=\frac{(2\pi)^{4}\delta(p_{a}+p_{b}-p_{c}-p_{d})}{2E_{a}2E_{b}2E_{c}2E_{d}}\mid \mathcal{M}\mid^{2}
\end{equation}
with transition amplitude $\mid \mathcal{M}\mid$. In the center of mass frame  Eq. (~\ref{tdrext}) can be simplified as
 \begin{equation}
\tau^{-1}(E_{a})=\sum_{b}\int\frac{d^{3}p_{b}}{(2\pi)^{3}}\sigma_{ab}\frac{\sqrt{s-4m^{2}}}{2E_{a}2E_{b}}f_{b}^{0}\nonumber
\end{equation}
\begin{equation}
\equiv\sum_{b}\int\frac{d^{3}p_{b}}{(2\pi)^{3}}\sigma_{ab}v_{ab}f_{b}^{0}
\label{relx}
\end{equation}
where $v_{ab}$ is relative velocity,  $\sigma_{ab}$ is the total scattering cross section, and $\sqrt{s}$ is center of mass energy.

For simplicity, the $\tau (E_{a})$ can be approximated to averaged relaxation time ($\tilde \tau$), which one can obtain by averaging Eq. (\ref{relx}) over the distribution function, $f_{a}^{0}$ as,

\begin{align}
    \label{relx1}
    \begin{split}
    {\tilde\tau}_{a}^{-1} & = \frac{\int\frac{d^{3}p_{a}}{(2\pi)^{3}}\tau^{-1}(E_{a})f_{a}^{0}}{\int\frac{d^{3}p_{a}}{(2\pi)^{3}}f_{a}^{0}}\\
    &=\sum_{b}\frac{\int\frac{d^{3}p_{a}}{(2\pi)^{3}}\frac{d^{3}p_{b}}{(2\pi)^{3}}\sigma_{ab}v_{ab}f_{a}^{0}f_{b}^{0}}{\int\frac{d^{3}p_{a}}{(2\pi)^{3}}f_{a}^{0}}\\
    &=\sum_{b}n_{b}\langle\sigma_{ab}v_{ab}\rangle .\end{split}
\end{align}
For a single component of pion gas, the above equation can be written as,
\begin{equation}
    \tilde\tau_{a}^{-1} = n_{b}\langle\sigma_{ab}v_{ab}\rangle,
\end{equation}

where $n_{b}=\int\frac{d^{3}p_{b}}{(2\pi)^{3}}f_{b}^{0}$ is the number density of pions.

In this work, for a pion gas, the equilibrium Bose-Einstein distribution (in the local rest frame) will be used, which is given by
 \begin{equation}
f_{a}^{0}= \Bigg[exp\bigg(\frac{E_{a}-\mu}{T}\bigg) -1\Bigg]^{-1}
\label{MB}
\end{equation}

 $\langle\sigma v\rangle$ is the thermal average cross-section multiplied by the relative velocity. Here, the scattering particles are considered hard spheres having a constant cross-section, $\sigma$.  Thus, $\langle\sigma v\rangle$ can be calculated as follows~\cite{Cannoni,Gondolo},
 \begin{equation}
 \langle\sigma_{ab} v_{ab}\rangle=\frac{\sigma \int d^{3}p_{a}d^{3}p_{b} v_{ab}f_{a}^{0}f_{b}^{0}}{\int d^{3}p_{a}d^{3}p_{b}f_{a}^{0}f_{b}^{0}}
 \label{thermalave}
 \end{equation}

The momentum space volume elements can be written as,
\begin{equation}
\label{eq16}
d^{3}p_{a}d^{3}p_{b}=8\pi^{2}p_{a}p_{b}dE_{a}dE_{b}~dcos\theta.
\end{equation}

On solving, Eq. ~\ref{thermalave} can be further written as,
\begin{widetext}
\begin{equation}
\label{eq17}
\langle\sigma_{ab}v_{ab} \rangle=\frac{\sigma\int_{}^{}8\pi^{2}p_{a}p_{b}dE_{a}dE_{b}~dcos\theta f_{a}^{0}f_{b}^{0}\frac{\sqrt{(E_aE_b-p_{a}p_{b}cos\theta)^{2}-(m_{a}m_{b})^{2}}}{{E_{a}E{_b}-p_{a}p_{b}cos\theta}}     }{\int_{}^{}8\pi^{2}p_{a}p_{b}dE_{a}dE_{b}~dcos\theta f_{a}^{0}f_{b}^{0}}.
\end{equation}
\end{widetext}
$\sigma$ is the hadronic collision cross-section. A constant value of $\sigma$ = 11.3 mb \cite{Tiwari:2017aon,Kadam:2015xsa} is used in the calculations, when the radius of pions is taken to be 0.3 fm.


Now, we move towards the calculation of viscosities in a pion gas which is under BEC. In such a situation, two types of collisions will occur: collisions between two excited particles and collisions between the excited particles and condensate particles. Thus, the reciprocal of relaxation time can be modified as the sum of the contributions from the above two types of collisions ($\tilde\tau_{12}^{-1} + \tilde\tau_{22}^{-1}$) \cite{ShazLowE}.

So, if $n_1$ is the density of particles in the condensed state and $n_2$ is that of in the excited state, the modified average relaxation time can be written as,
\begin{widetext}
 \begin{equation}
{\tilde\tau}^{-1}= \bigg[ n_{1}\langle\sigma_{12}v_{12}\rangle + n_{2}\langle\sigma_{22^{'}}v_{22^{'}}\rangle \bigg],
\end{equation}

\begin{equation}
\label{eq17}
\langle\sigma_{12}v_{12} \rangle= \frac{\sigma\int_{}{}d^{3}p_{1}d^{3}p_{2}v_{12}f_{1}^{0}f_{2}^{0}}{\int_{}{}d^{3}p_{1}d^{3}p_{2}f_{1}^{0}f_{2}^{0}}.
\end{equation}

\begin{equation}
\label{eq17}
\langle\sigma_{22^{'}}v_{22^{'}} \rangle=\frac{\sigma\int_{}^{}8\pi^{2}p_{2}p_{2^{'}}dE_{2}dE_{2^{'}}~dcos\theta f_{2}^{0}f_{2^{'}}^{0}\frac{\sqrt{(E_{2}E_{2^{'}}-p_{2}p_{2^{'}}cos\theta)^{2}-(m_{2}m_{2^{'}})^{2}}}{{E_{2}E_{2^{'}}-p_{2}p_{2^{'}}cos\theta}}     }{\int_{}^{}8\pi^{2}p_{2}p_{2^{'}}dE_{2}dE_{2^{'}}~dcos\theta f_{2}^{0}f_{2^{'}}^{0}}.
\end{equation}
\end{widetext}

In this work, we estimate the shear and bulk viscosities using a kinetic approach with the relaxation-time approximation (RTA) \cite{Reif1965}. The RTA is based on an ansatz for the collision integral in the Boltzmann equation, and its accuracy depends crucially on this ansatz and the relaxation time considered \cite{Plumari:2012ep}. There are several other methods, such as the Chapman-Enskog (CE) approximation \cite{Groot1980} that can be used for the estimation of transport coefficients. The CE method allows one to obtain solutions with an accuracy that depends upon the order of approximation used \cite{Plumari:2012ep, Wiranata:2012br}. An alternative approach is based on the Kubo formalism \cite{Huang:2011dc}, where the non-equilibrium processes are regarded as the response of the system to an external perturbation and its computation involves the correlation functions. Comparative studies~\cite{Plumari:2012ep, Wiranata:2012vv, Demir:2014kda} have shown that the RTA can give estimates that differ numerically from Green-Kubo or CE results, while CE expansions typically improve agreement with Green-Kubo estimations. While Kubo formalism \cite{Kadam:2014cua, Karsch:2007jc, Kadam:2014xka, Ghosh:2014ija} is more fundamental and systematically incorporates quantum statistical effects, the RTA \cite{Chakraborty:2010fr, Khvorostukhin:2010aj, Ghosh:2013cba} is a simplified model of the full collision integral and is used for parametric studies and systems where a clear particle-like description exists.

In heavy-ion collisions, the speed of sound is studied to understand the hydrodynamical evolution of the matter created in such collisions. It can also be used to probe the degrees of freedom as it is related to the equation of state of the system \cite{Khuntia:2016ikm}. In thermodynamics, the speed of sound in matter is expressed as the change in pressure with energy density at constant entropy density ($s$) per number density ($n$), i.e., $s/n$, and is given by
\begin{equation}
\label{cs2}
c_{\rm s}^{2} (T,\mu) = \frac{\partial P}{\partial \varepsilon}\bigg\vert_{s/n},
\end{equation}
where the pressure $P$ and the energy density $\varepsilon$ for a pion gas in BEC can be written in a similar form to Eq.~(\ref{n_finite}) as
\begin{align}
\label{prs}
P = 
-\frac{Tg}{2\pi^{2}}\int_{}^{}p^{2}dp \ln{\bigg[1-\rm exp\bigg(-\frac{E-\mu}{T}\bigg)\bigg]},
\end{align}

and

\begin{equation}
\label{energy}
\varepsilon =  \frac {gm}{V [{\rm exp} (\frac{m - \mu}{T}) - 1]} + \int \frac{d^3p}{(2\pi)^3} \frac {gE}{{\rm exp} \bigg(\frac{\sqrt{{\bf p}^2 + m^2} - \mu}{T}\bigg) - 1}
\end{equation}

Eq.~(\ref{cs2}) can be further simplified as,

\begin{equation}
\label{eq19}
    c_{\rm s}^2 (T,\mu) = \frac{\frac{\partial P}{\partial T} + \frac{\partial P}{\partial \mu}\frac{d\mu}{dT}}{\frac{\partial \varepsilon}{\partial T} + \frac{\partial \varepsilon}{\partial \mu}\frac{d\mu}{dT}},
\end{equation}
where,

\begin{equation}
\label{eq20}
    \frac{d\mu}{dT} = \frac{s\frac{\partial n}{\partial T} - n\frac{\partial s}{\partial T}}{n\frac{\partial s}{\partial \mu} - s\frac{\partial n}{\partial \mu}}.
\end{equation}
\begin{figure*}[ht!]
\begin{center}
\includegraphics[scale = 0.44]{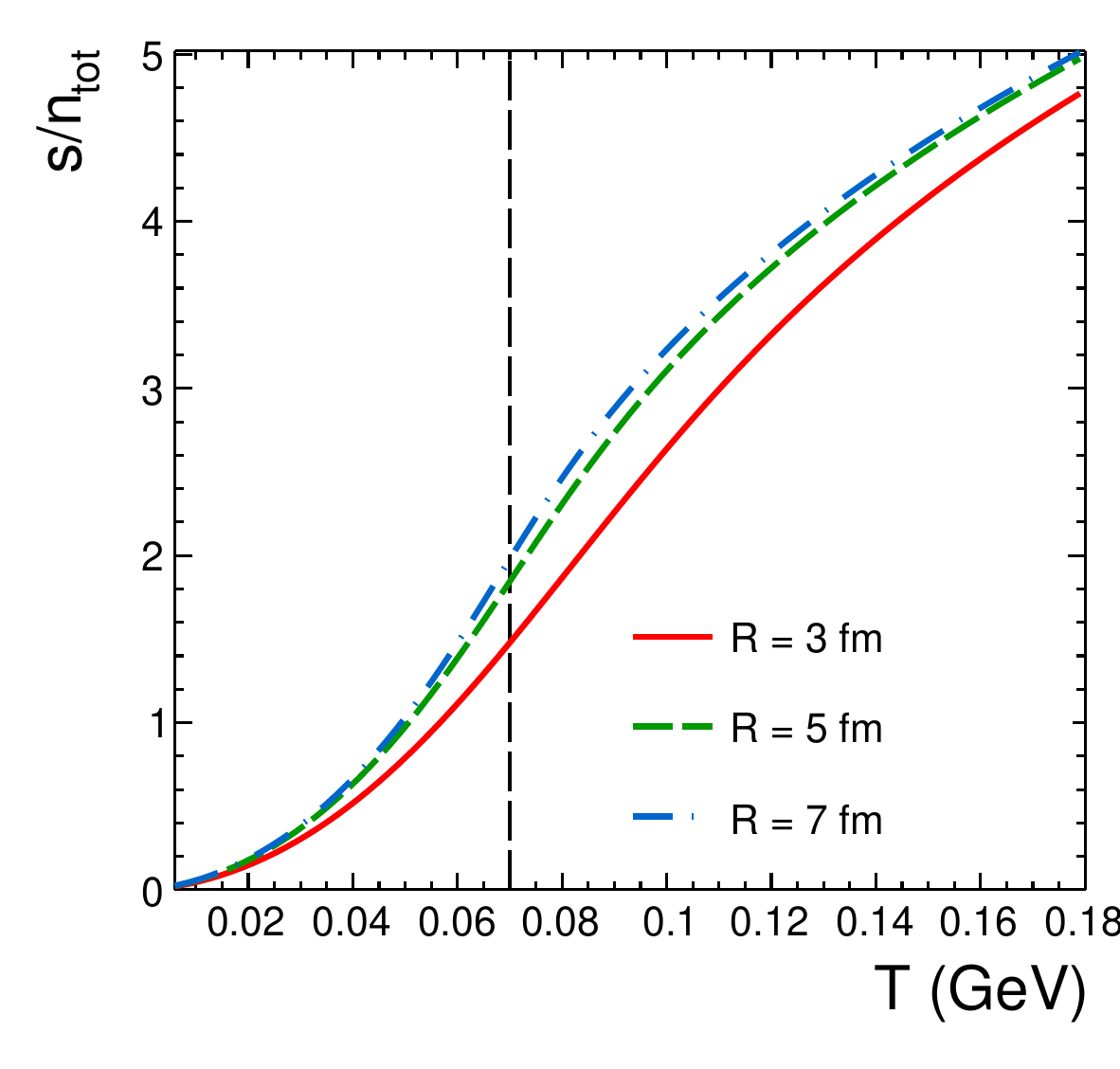}
\includegraphics[scale = 0.44]{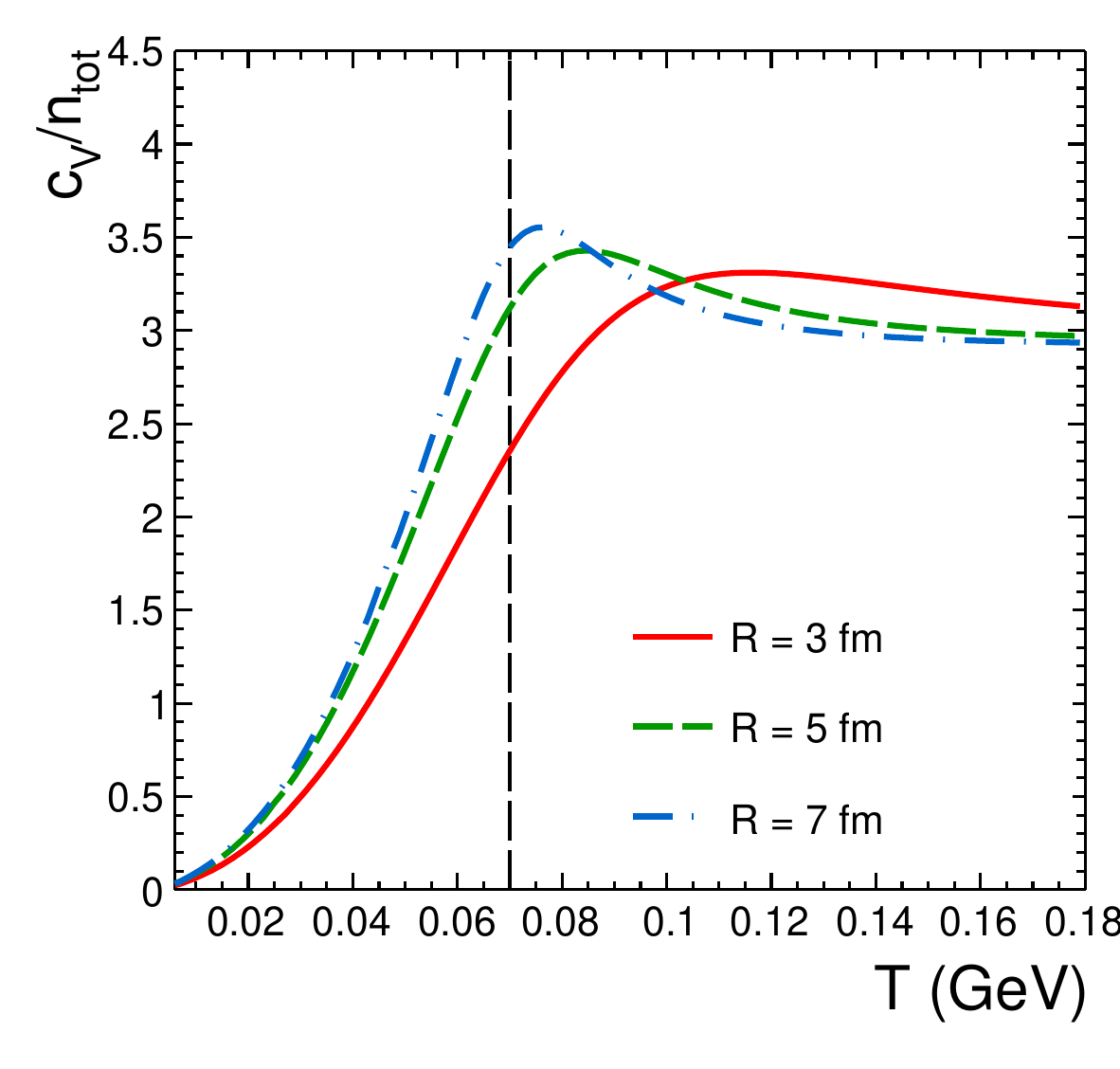}
\caption{The scaled entropy density, $s/n_{\rm tot}$ (left panel) and specific heat, $c_v/n_{\rm tot}$ (right panel) as a function of temperature for various system sizes with $R$ = 3, 5, and 7 fm.}
\label{fig3}
\end{center}
\end{figure*}
Moreover, the entropy density and specific heat ($c_v$) in a BEC can be obtained as,
\begin{equation}
    \label{entropy}
    \begin{split}
    s=
    &-\frac{g}{2\pi^2} \int_{0}^{\infty} p^2 dp \Big[\ln\{1 - \exp[-(E-\mu)/T]\}\\ 
    &\ \ \ \ - \frac{T}{\exp[(E-\mu)/T] - 1}\Big(\frac{E-\mu+\mu^\prime T}{T^2}\Big)\Big],
    \end{split}
\end{equation}

\begin{equation}
    \begin{split}
        \label{eqcv}
        c_v =& \frac {g}{V} \frac{m~\exp[(m - \mu)/T]}{\big(\exp[(m - \mu)/T] - 1\big)^2} \Big(\frac{m-\mu+\mu^\prime T}{T^2}\Big)\\
        +&\frac{g}{2\pi^2}\int p^2dp \Bigg[\frac{E~\exp[(\sqrt{p^2+m^2}-\mu)/T]}{\big(\exp[(\sqrt{p^2+m^2}-\mu)/T] - 1\big)^2}\\
        &\ \ \ \ \ \ \ \ \ \ \ \ \ \ \ \ \ \ \ \ \ \times \Big(\frac{E-\mu+\mu^\prime T}{T^2}\Big)\Bigg],
    \end{split}
\end{equation}
where $\mu^\prime = \partial\mu/\partial T$.
By substituting the above equations in Eq.~(\ref{cs2}), we can estimate the speed of sound of a pion gas under BEC.

\section{Results and Discussion}
\label{res}

Before discussing the viscosity of the system, we first estimate the pion condensate fraction for different cases of system size using Eq.~(\ref{cfraction}). This is shown in Fig.~\ref{fig1} along with the temperature dependence of the chemical potential. As already discussed, the critical temperature $T_c$ depends upon the total number density of the system considered, and therefore the BEC phenomenon can happen even earlier (at high $T$, as one can see from the red curve in Fig.~\ref{fig2}). In addition to this, external factors may lead to a larger number of particles in the condensate and hence achieve the BEC transition early. For example, the magnetic field and rotation can affect the condensation of the pion gas. Interestingly, when we introduce rotation to the system, we introduce a different type of chemical potential, with the term $\Omega$ acting as the rotational chemical potential. Similarly, we can also notice from the fundamental Euler's thermodynamic relation that introducing an external magnetic field to the system introduces a magnetic chemical potential. Thus, we can easily infer that any addition of chemical potential will increase the number of particles in the condensate. In Ref.~\cite{Siri:2024cjw}, the impact of rotation is studied on the thermodynamic properties of a BEC as well as on the BEC transition temperature along with the condensate fraction. The authors showed that a rigid rotation with small angular velocity decreases the speed of sound as well as the BEC transition temperature in comparison to the non-rotating BE gas. In the non-relativistic limit, they found that the rotation is responsible for a discontinuous specific heat at the transition temperature, though it shows a continuous behaviour for the non-rotating case at the corresponding transition temperature. This indicates that the rotation may affect the features of the BEC transition of a pion gas. Another important and physically relevant effect often present in such collisions is the presence of strong magnetic fields. The authors in Ref.~\cite{Ayala:2016awt} demonstrated that the BEC transition temperature increases in the presence of a magnetic field. They have shown that the critical chemical potential for a given temperature decreases from the value it had in the absence of a magnetic field, enhancing the condensation. A study on the charged pion condensation under the effect of rotation at an external constant magnetic field \cite{Liu:2017spl} shows that with an increase in rotation, the pion condensation also increases. The inclusion of a magnetic field shifts the energy levels of charged particles via Landau quantization, and the lowest Landau level (LLL) can facilitate condensation, in particular, the splitting down of the LLL of $\pi^+$ pion causes its condensation. While the current work focuses on thermal BEC in a pion gas without external fields, incorporating these rotational and magnetic effects may alter the critical temperature for BEC and also affect the transport coefficients. A full treatment of these coupled effects represents an important direction for future exploration.
\begin{figure*}
    \centering
    \includegraphics[scale = 0.42]{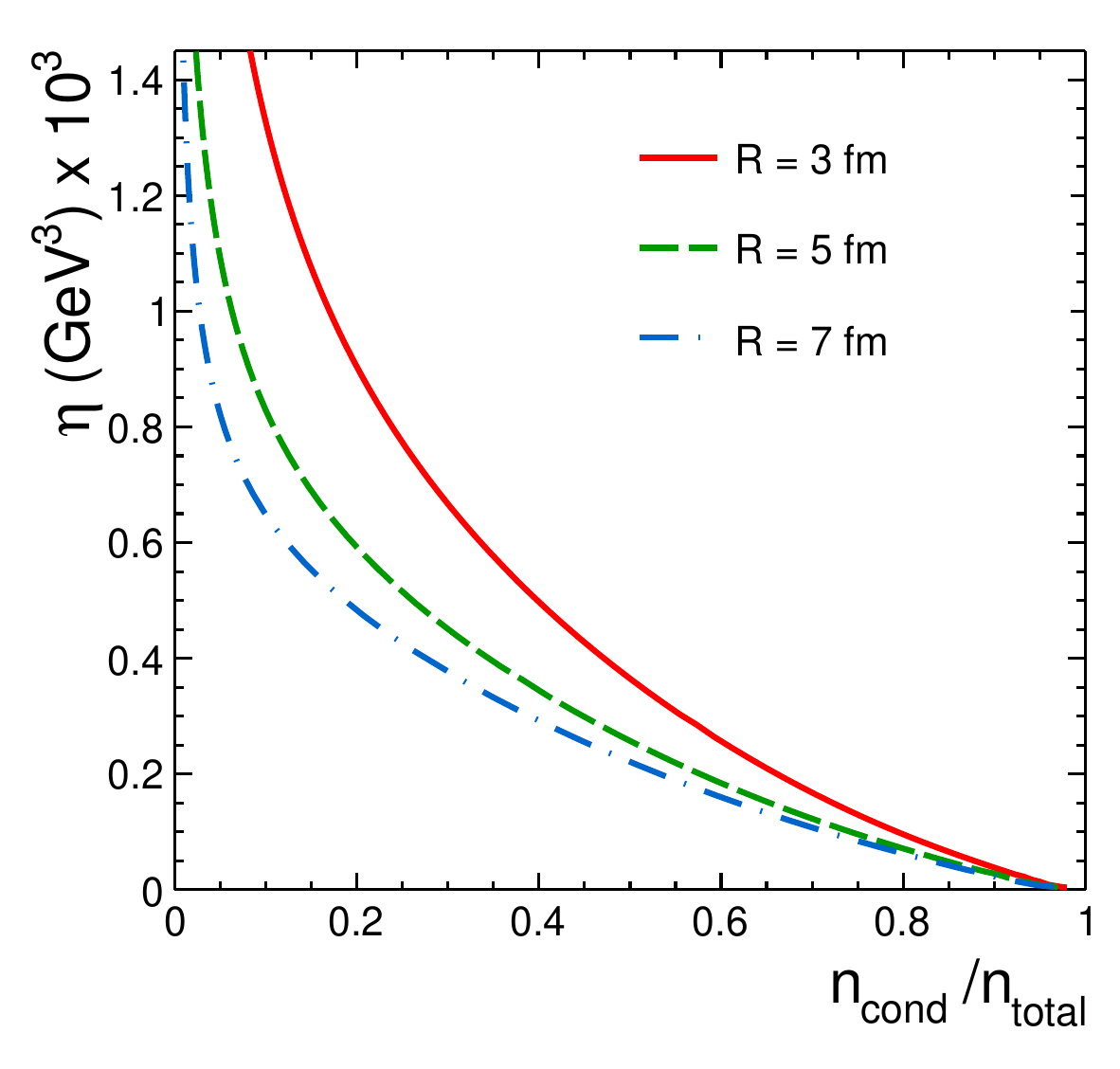}
    \includegraphics[scale = 0.42]{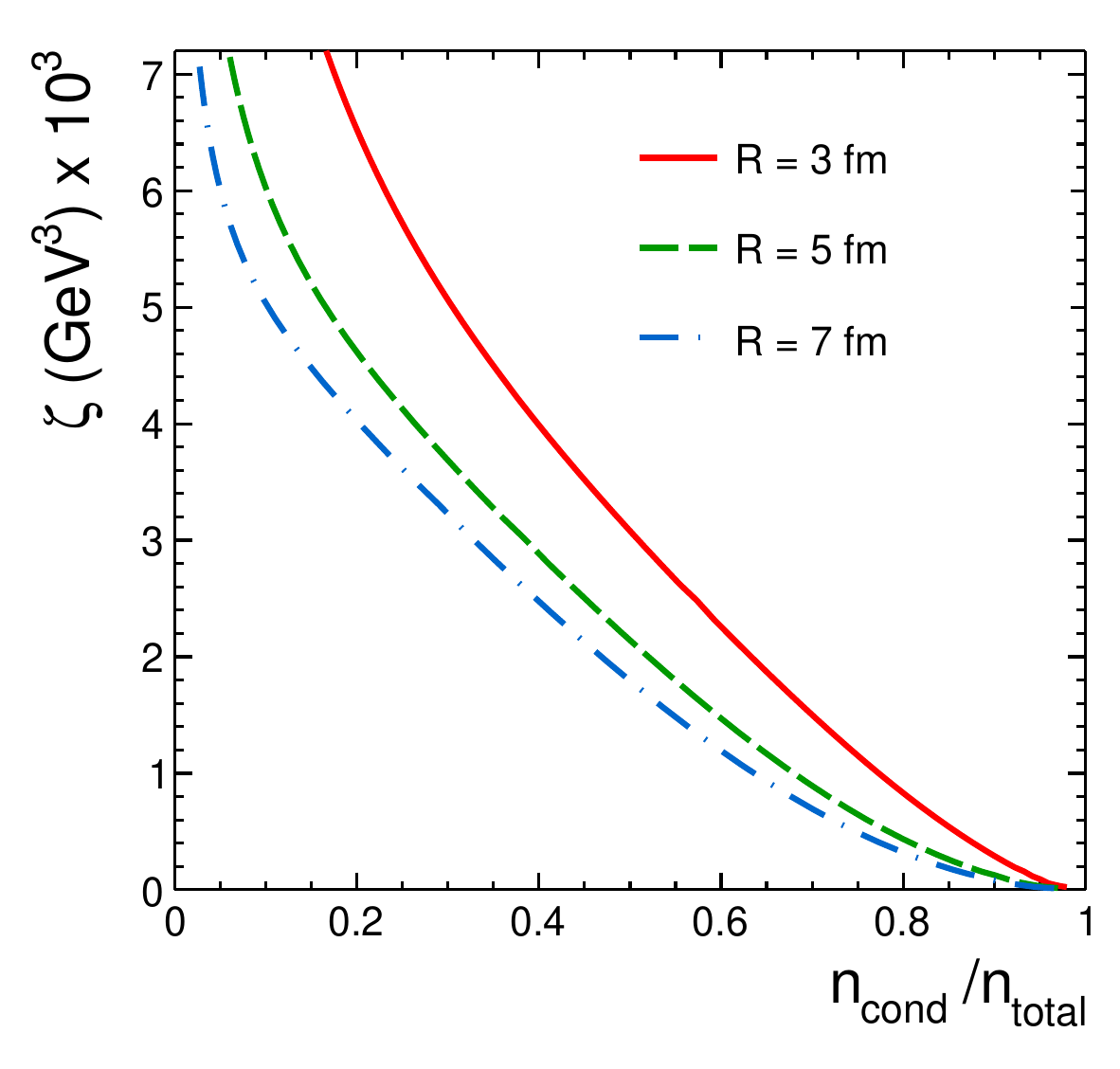}
    \caption{Shear viscosity (left panel) and bulk viscosity (right panel) as functions of the ratio of the number of pions in the condensate to the total number of pions for various system sizes with $R$ = 3, 5, and 7 fm.}
    \label{fig4}
\end{figure*}

\begin{figure*}
    \centering
    \includegraphics[scale = 0.42]{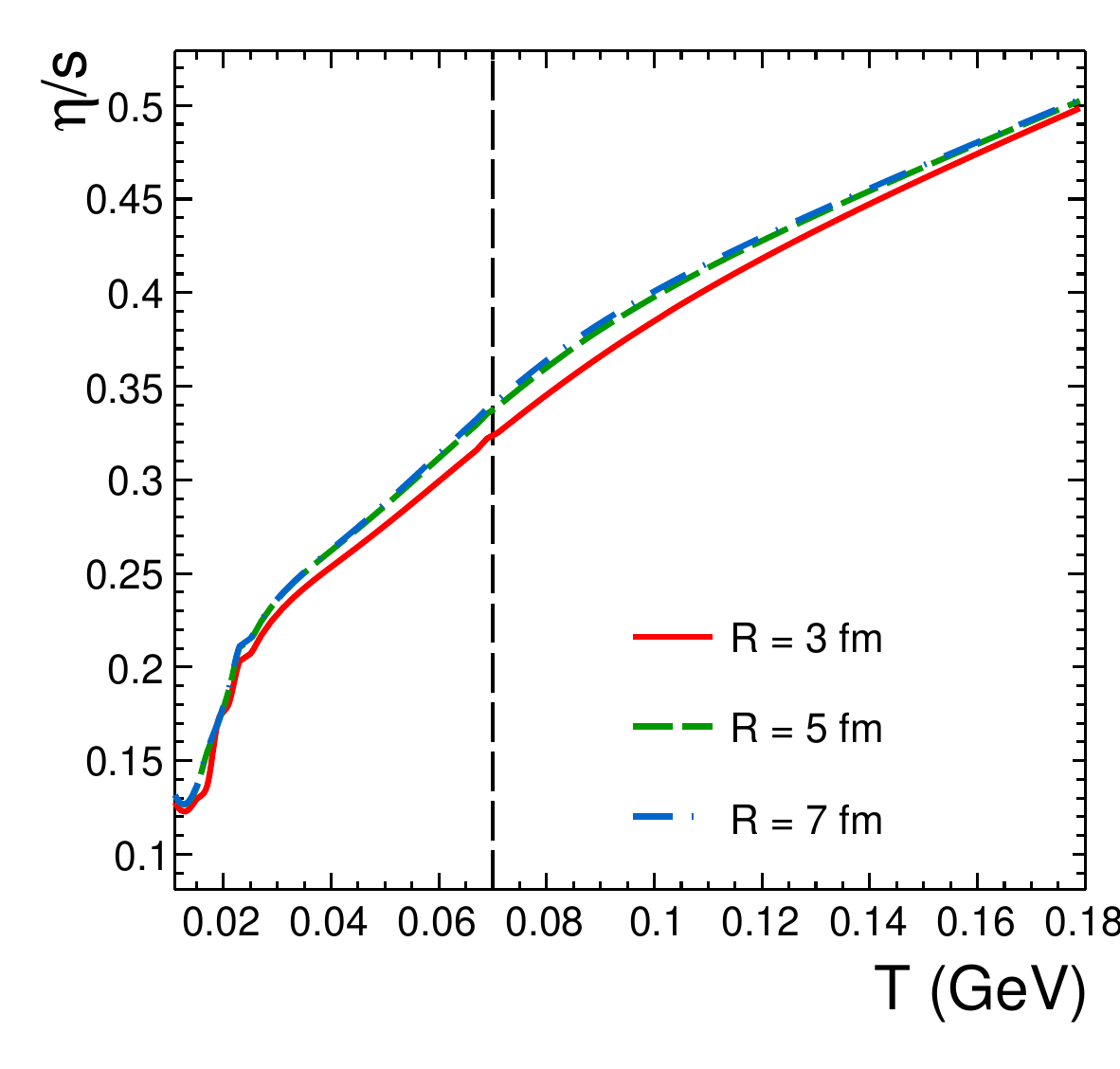}
    \includegraphics[scale = 0.42]{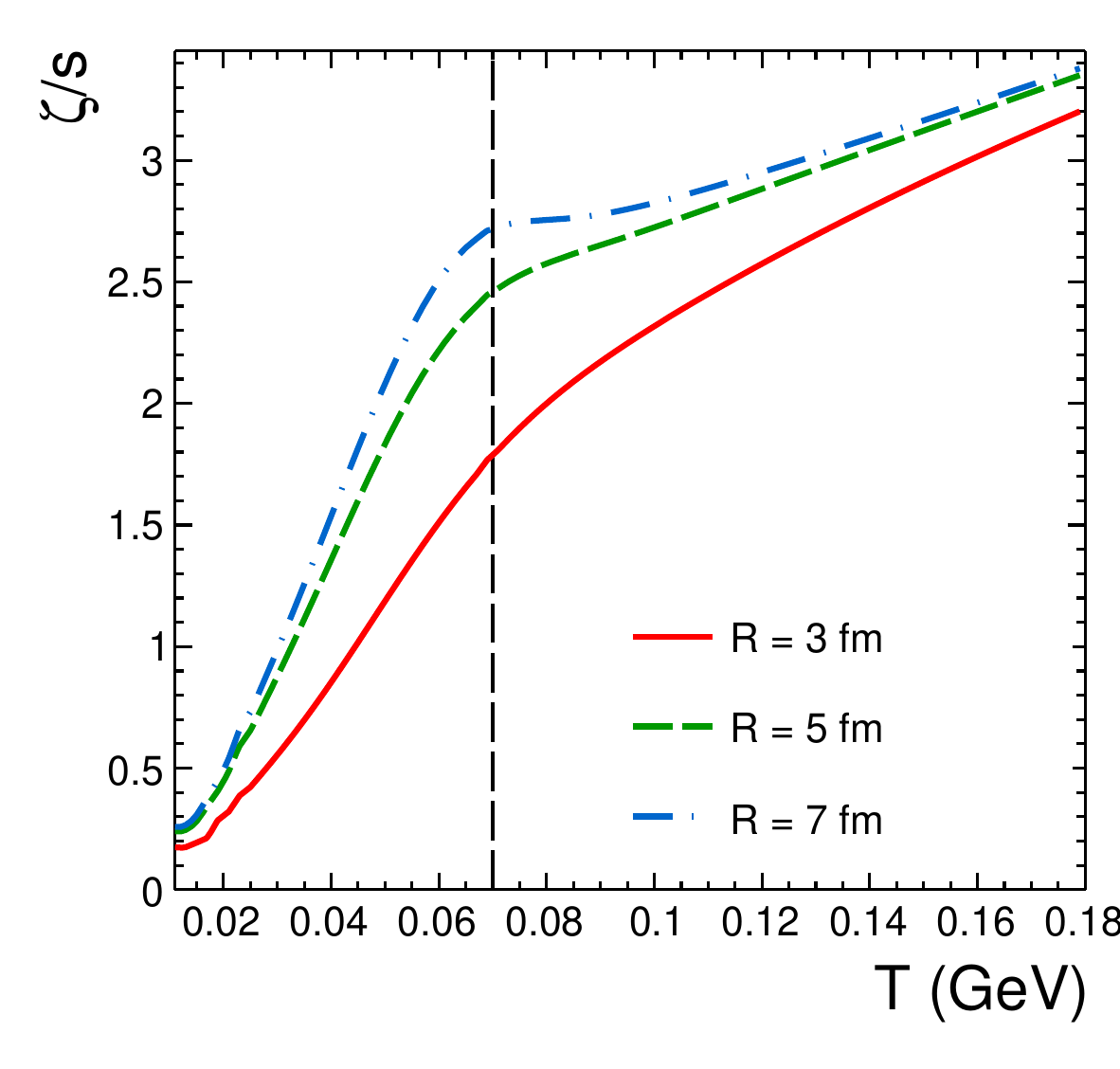}
    \caption{Shear viscosity to entropy density ratio (left panel) and bulk viscosity to entropy density ratio (right panel) of a pion gas under BEC as functions of temperature for various system sizes with $R$ = 3, 5, and 7 fm.}
    \label{fig5}
\end{figure*}
\begin{figure*}
    \centering
    \includegraphics[scale = 0.42]{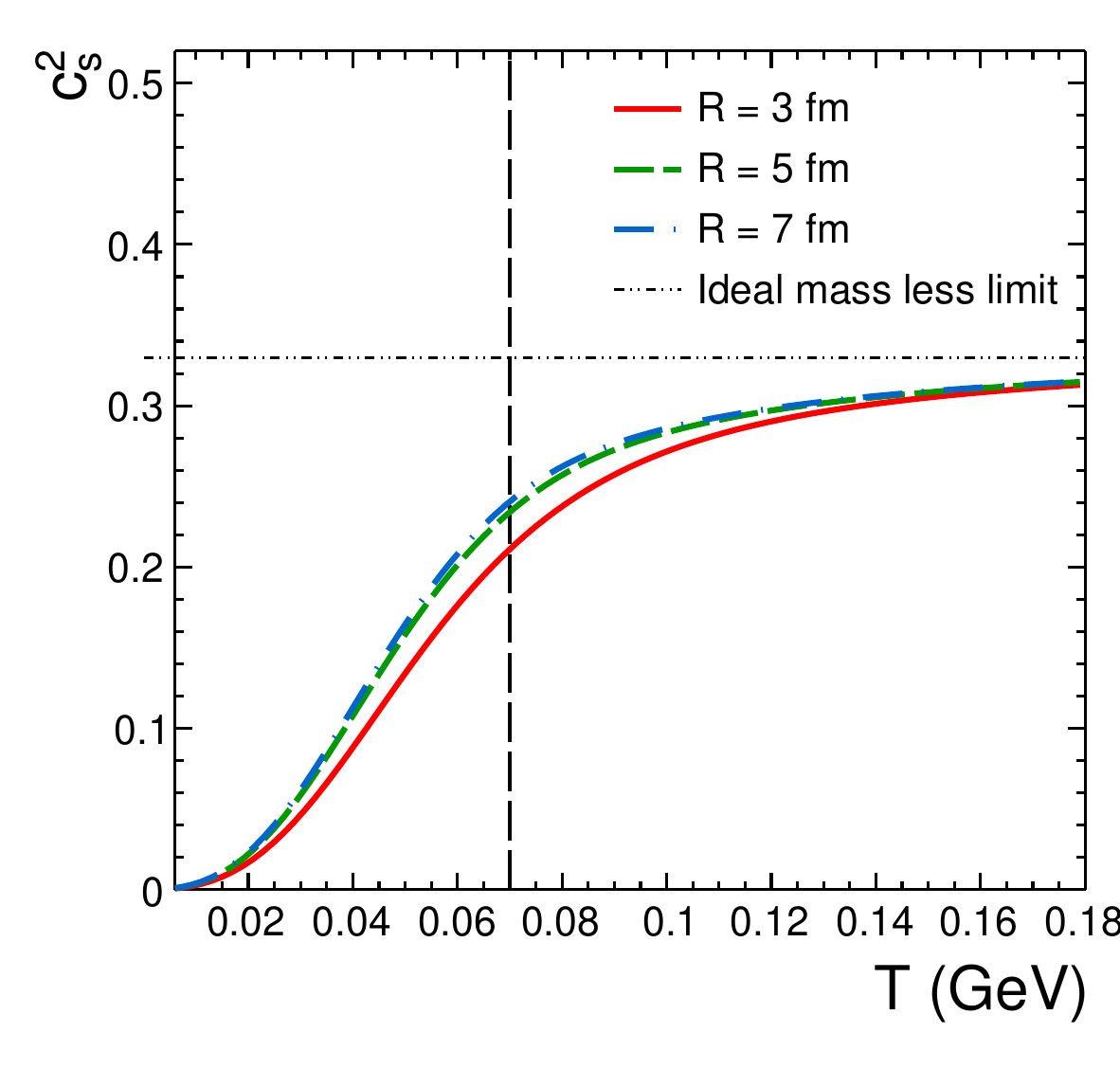}
    \includegraphics[scale = 0.42]{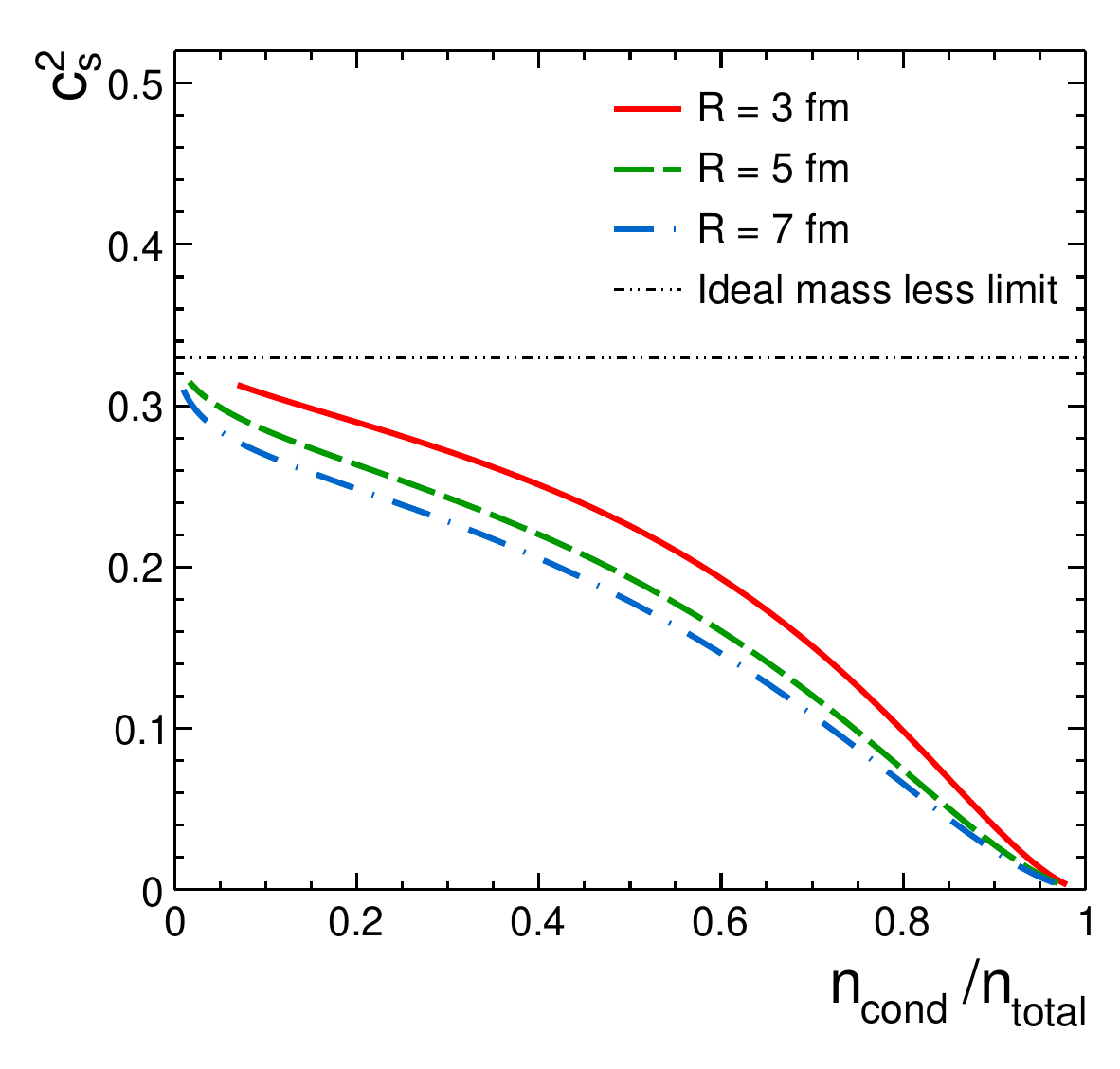}
    \caption{Speed of sound as a function of temperature (left panel) and as a function of ratio of number of pions in the condensate to the total number of pions (right panel) for system sizes with $R$ = 3, 5, and 7 fm.}
    \label{fig6}
\end{figure*}
Phase transitions are usually identified by a discontinuous behaviour of some thermodynamic quantities. A first-order phase transition, like a liquid-gas transition, involves a discontinuity in the first-order derivative of free energy. For example, the entropy density is discontinuous at $T_c$. The interparticle interactions are responsible for such phase transitions. In an ideal Bose gas, the BEC transition, however, is a quantum-statistical phase transition that occurs entirely due to the Bose-Einstein statistics. This transition is characterized by the sudden increase in the ground state population below $T_c$. The left panel of Fig.~\ref{fig3} shows the variation of entropy density scaled by total number density as a function of temperature. In our calculations, we estimate various thermodynamic and transport coefficients for a range of temperatures starting from 5 MeV to 180 MeV. One can observe that for all the considered cases of system size, the entropy density is very low in the condensed state and starts increasing with temperature. There is a continuous transition at $T_c$ as discussed. The entropy density estimated here corresponds solely to the excited states. In Ref.~\cite{Kim2018}, the entropy of an ideal BEC in a three-dimensional trap was found to exhibit similar behaviour. Though the entropy of the ground state is non-zero, it is shown that the total entropy is equal to the entropy of the excited state. The entropy due to the correlation between the ground state and excited state, arising from the constraint of fixed density, cancels the entropy of the ground state. Nevertheless, the change in entropy is continuous at $T_c$ with a change in slope of the curve. The same behaviour is already observed experimentally in a two-dimensional trapped photon gas \cite{nature_com2016}, where it matches with the expected theoretical ideal Bose gas behaviour. The entropy per photon decreases with $T/T_c$ in accordance with the third law of thermodynamics. The entropy per particle is also observed for a two-dimensional atomic gas in experiments \cite{Yefsah2011}. The key point to be noted is that a relativistic hot pion gas undergoing a BEC transition has similar thermodynamic properties to those of a low-energy BEC, strongly supporting the universal properties of BEC. 

Another interesting quantity to study regarding the properties of phase transitions is the specific heat per particle. For a Bose gas, it is observed that $c_v$ increases with temperature, reaches a maximum, and then gradually decreases with $T$. A characteristic $\Lambda$-like shape (also leads to the name $\Lambda$-transition as in the case of liquid helium~\cite{Thorn2002}) is observed for $c_v$ as a function of $T$. A cusp singularity is observed at $T_c$ in both theoretical \cite{pethick2008, Pathria2011} and experimental observations~\cite{nature_com2016}. A similar behaviour is obtained for $c_v$ for a relativistic pion gas in the thermodynamic limit \cite{Begun:2008hq}. It is observed that $c_v$ approaches 3/2 at high $T$ (for $T>>T_c$) for a non-relativistic Bose gas \cite{Pathria2011}, whereas it saturates at $\sim3$ for the relativistic pion gas~\cite{Begun:2008hq} as shown in Fig.~\ref{fig3}. As the volume increases, the maximum is more pronounced near the $T_c$, and in the thermodynamic limit, it must show a cusp-like behaviour as shown in Ref.~\cite{Begun:2008hq}. By definition, the pion gas considered here exhibits a continuous (second-order) phase transition to a condensed state with a smooth evolution of thermodynamic quantities like entropy density without any involvement of latent heat~\cite{Landau1980}. However, this transition lacks conventional second-order criticality, such as diverging correlation lengths or higher-order susceptibility, as there is no mechanism in an ideal Bose gas to mediate long-range correlations. In contrast, introducing interactions among the pions will lead to a phase transition that depends upon the nature of the interaction. In fact, a first-order Bose-Einstein condensation is predicted for harmonically trapped bosons with both attractive and repulsive interactions \cite{HuiHu2021} and for a two-component Bose mixture with repulsive interactions \cite{Jakub2024}. One can also refer to \cite{Olivares2010} for a discussion on the order of the BEC transition in a Bose gas predicted using mean field theory. The interaction can also influence the BEC properties of the pion gas. In Ref.~\cite{Savchuk:2020yxc}, the authors investigate how various types of repulsive interactions affect both the BEC transition temperature and particle number fluctuations. It is observed that the characteristic BEC transition line varies across different interacting models like excluded volume (EV), mean field (MF) and effective mass (EM). Therefore, the onset of BEC in the presence of interactions happens at different temperatures. Furthermore, the behavior of particle number fluctuations differs between the EV model and the ideal pion gas, where fluctuations diverge below $T_{c}$. On the other hand, it remains finite and continuous in EM and MF models. Thus, examining the impact of interactions on the thermodynamic and transport properties of a gas in BEC is worthwhile and essential. 



Now, to understand the viscous properties of the pion gas, the Boltzmann transport equation is used under relaxation time approximation. One can also generalize this description using a quasiparticle framework, where the degrees of freedom for the transport phenomena include the medium-modified thermal excitations with effective masses. It is to be noted that here, the Boltzmann equation in the RTA primarily describes the dynamics of thermal excitations and assumes a gas of hadrons with well-defined masses and momenta. However, below the transition temperature, the dynamics of the condensate are better described by the Gross-Pitaevskii equation (GPE)~\cite{ShazLowE}, which captures the coherent evolution of the macroscopic wavefunction in a fully condensed state. Moreover, a more complete description would involve coupled equations for the condensate and non-condensate interactions, such as the generalized Gross-Pitaevskii and quantum kinetic equations, as discussed in ~\cite{ShazLowE}.  Including these effects would require a more elaborate two-fluid formalism, and will be discussed elsewhere. In this work, we estimate the shear and bulk viscosities of a pion gas under BEC using the Boltzmann transport equation under RTA. This becomes a little tricky. As already discussed, under BEC, the pion gas will have two kinds of subsystems: excited (non-condensate) pions with finite momentum and pions at the zero or near-zero momentum ground state with relative momentum zero. Due to this, two types of collisions occur in the system: one between two excited-state pions and one between an excited-state pion and a pion in the BE condensate. Thus, the conventional relaxation time approximation formalism has to be modified accordingly. With this assumption, the shear and bulk viscosity as functions of the ratio $n_{\rm cond}/n_{\rm tot}$ for various system sizes are plotted in Fig.~\ref{fig4}. This is helpful to understand how the viscosity changes if the system has most of its pions in the condensate. While shear viscosity resists any change to the shape of a fluid element, the bulk viscosity acts against any change in the fluid's volume through expansion or compression. Along with the shear viscosity, the bulk viscosity can affect the flow by anisotropic expansion and compression of the matter produced in heavy-ion collisions \cite{Denicol:2009am, Sahu:2020mzo}. As one observes here, the system's viscosity is higher if most of the pions are in the excited states. On the other hand, if most of the pions in the system are in the condensate, the system's viscosity is lower; it approaches zero, suggesting that a system with all the pions in the condensate becomes inviscid. This aligns with the finding of low-temperature cold helium atom BEC, which is generally regarded as a superfluid due to its extremely low viscosity. Moreover, it is interesting to note that a system size dependency can also be observed. The decreasing trend of $\eta$ and $\zeta$ is steeper for higher system size; however, the trend for all system sizes becomes the same at a higher ratio of $n_{\rm cond}/n_{\rm tot}$. While studying system-size dependency, it is crucial to check the hydrodynamic applicability to the system under consideration. One can estimate quantities like mean free path ($\lambda$) and Knudsen number ($Kn$) \cite{Scaria:2022yrz} to validate the use of hydrodynamics ($Kn\ll1$). In this work, for a total pion density of 0.1 fm$^{-3}$, systems with radii $R$ = 7 and 5 fm satisfy the criterion. For $R$ = 3 fm, one obtains $Kn\sim1$ and is considered here only to study the system dependencies of different quantities. 

In Fig.~\ref{fig5}, we have plotted the shear viscosity to entropy density ratio and bulk viscosity to entropy density ratio as functions of temperature for various system sizes. For a system with a smaller radius, both viscosities are significantly lower in the low-temperature regime where the condensed fraction is large. However, at a higher temperature regime, a large contribution to the viscosities comes from the excited pions. Also, the system-size dependency of these viscosities becomes less as one goes towards the high $T$ region. A system with a smaller radius will have a higher density and, thus, a higher number of interactions, resulting in higher $\eta$ as shown in Fig.~\ref{fig4}. But, the entropy density is even higher as in Fig.~\ref{fig3}. Therefore, owing to the number of excited pions, the ratio $\eta/s$ and $\zeta/s$ in Fig.~\ref{fig5} is small at low temperature and increases with increasing $T$. Considering interactions along with a more generalized BTE and GPE equations for the Bose gas may possibly modify the results. Similar to shear, the bulk viscosity is also sensitive to the ideal gas approximation. Various QCD-inspired models predict a peak for the $\zeta/s$ at the QCD critical temperature due to large deviation from conformality and decreases again to the hadronic phase at low temperature \cite{Song:2008hj}. Though it is observed that near the BEC transition, there is an enhancement in the trace anomaly, which in turn increases the $\zeta/s$, the sharp decrease at low temperature requires a thorough investigation and shows the need to consider interactions for a proper understanding of the dissipative properties in the condensate phase at temperatures close to zero. For all the system sizes, the trend of both $\eta/s$ and $\zeta/s$ increases as a function of temperature, although the value of $\zeta/s$ is higher in magnitude than the value of $\eta/s$. We observe that BEC affects the viscosity of the system considerably. The temperature dependence of these quantities presents a stark contrast to the behavior of a conventional, non-condensed pion gas. Here, we study the transport properties of a BEC considering a fixed pion density. This implies a temperature dependence of the chemical potential, which decreases with temperature, playing a major role in the observed trend of the viscosity to entropy ratio. In the absence of a condensate, the viscosities are determined entirely by thermal excitations of a pion gas. For example, in Ref.~\cite{Ghosh:2014qba}, the authors have estimated the shear viscosity to entropy density ratio of pion gas at zero chemical potential, where it is observed to decrease with temperature close to the KSS limit and is comparable to those also obtained in Ref.~\cite{Lang:2012tt, Fernandez-Fraile:2006kxe, Kalikotay:2024yeh}. The bulk viscosity to entropy density ratio for a pion gas with physical pion mass is found to decrease with temperature below the QCD crossover temperature \cite{Lu:2011df}, although for a massless pion gas, it reaches a maximum near the crossover temperature \cite{Chen:2007kx}. In Ref.~\cite{Mitra:2013gya}, the authors have calculated the bulk and shear viscosities of a pion gas by solving the relativistic transport equation in
the Chapman-Enskog approximation. Here, the effect of early chemical freeze-out in heavy ion collisions is implemented through a temperature-dependent pion chemical
potential, which affects their behaviour as a function of temperature. Both the shear and bulk viscosity to entropy ratio increase with $T$ for temperature-dependent chemical potential, in contrast to those at zero chemical potential, where they decrease with $T$.

Furthermore, the effect of BEC on the speed of sound in a pion gas is explored. The speed of sound is a very important observable that tells us how interactive the system is. In the left panel of Fig.~\ref{fig6}, we have plotted the squared speed of sound of the hot pion gas under BEC. 
On the right-hand side panel of fig.~\ref{fig6}, we have plotted $c_{\rm s}^{2}$ as a function of the ratio $n_{\rm cond}/n_{\rm tot}$ for various system sizes. Counterintuitively, the squared speed of sound $c_s^2$ decreases with increasing condensate population, despite a simultaneous drop in viscosities as shown in Fig.~\ref{fig6}. In typical fluids, lower viscosity often correlates with a stiffer equation of state and higher $c_s^2$, making this behavior unexpected. However, in a pion Bose-Einstein condensate (BEC), condensed pions do not contribute significantly to the pressure, while the energy density continues to rise. This leads to a softening of the equation of state, hence the observed reduction in $c_s^2$. This dual reduction, while seemingly counterintuitive from classical hydrodynamics perspectives, emerges naturally from the quantum statistics of the BEC phase: the viscosity suppression originates from macroscopic coherence where particles occupy the lowest momentum space, which suppresses momentum transfer between fluid layers (shear viscosity) and volume expansion resistance (bulk viscosity). As more pions enter the zero-momentum condensate state, the phase space for dissipative processes ($\pi_{\text{excited}}$-$\pi_{\text{excited}}$ and $\pi_{\text{excited}}$-$\pi_{\text{cond}}$ collisions) dramatically shrinks. On the other hand, the sound speed softening reflects the condensate's diminished contribution to pressure. This creates a unique quantum fluid regime where reduced viscosity results in a near-dissipationless flow while the softer EoS alters density perturbation propagation. This can have potential experimental signatures in the $p_T$-dependence of elliptic flow and freeze-out dynamics at LHC/RHIC energies. In particular, this reduction in pressure gradient and inviscid behavior may result in the suppression of the flow observable, like elliptic flow. An event-by-event variation of the azimuthal distribution, specifically at low $p_{\rm T}$, would be an interesting thing to see in experiments. Though the resonance decay populates the low $p_{\rm T}$ region of pion $p_{\rm T}$ spectra, high-purity pion measurements at low $p_{\rm T}$ may help to see any increase in its number. These effects are particularly sensitive to the finite-size modifications of the BEC critical temperature $T_c$.

\section{Summary}
\label{sum}

This article examines the effects of Bose-Einstein condensation on the macroscopic transport properties of a hot, dense pion gas, a system of particular relevance to the hadronic phase of matter created in ultra-relativistic heavy-ion collisions at RHIC and LHC energies. By solving the relativistic Boltzmann transport equation within the relaxation time approximation framework, the systematic calculations of three crucial physical quantities are performed: the shear viscosity ($\eta$), bulk viscosity ($\zeta$), and speed of sound ($c_s$) for a pion gas undergoing BEC transition. The study reveals several remarkable phenomena: 

\begin{itemize}
    \item Both shear and bulk viscosities exhibit dramatic suppression when pions begin to condense into the ground state, with this effect being particularly pronounced in systems with higher chemical potentials ($\mu \sim 120$ MeV) or smaller spatial extents ($R = 3-7$ fm).

    \item The viscosity coefficients show universal scaling with the condensate fraction $N_{\text{cond}}/N_{\text{total}}$, asymptotically approaching zero as this ratio tends toward unity - behavior strongly reminiscent of superfluid hydrodynamics observed in cold atomic BEC systems.

    \item Counterintuitively, while the viscosities decrease, the squared speed of sound $c_s^2$ also decreases with increasing condensate population, indicating a softening of the equation of state as more pions enter the BEC phase.
    
\end{itemize}   

This dual effect on transport properties (reduced viscosity but also reduced sound speed) suggests complex modifications to the hydrodynamic evolution of the hadronic medium. The finite-size effects are analysed, demonstrating that a larger collision system ($R = 7$ fm) shows more pronounced BEC signatures approaching the thermodynamic limit. These findings have potentially significant implications for interpreting collective flow measurements in heavy-ion collisions, particularly the temperature and system-size dependence of elliptic flow coefficients. This work provides fundamental insights into how quantum statistical effects can dramatically alter the dissipative properties and equation of state of strongly interacting matter under extreme conditions.

\section*{Acknowledgement}

KKP  acknowledges the doctoral fellowship from the UGC, Government of India. The authors gratefully acknowledge the DAE-DST, Govt. of India funding under the mega-science project -- “Indian participation in the ALICE experiment at CERN" bearing Project No. SR/MF/PS-02/2021-IITI (E-37123).

\end{document}